# Harnessing Unipolar Threshold Switches for Enhanced Rectification

Md Mazharul Islam[1], Shamiul Alam[1], *Graduate Student Member, IEEE,* Garrett S. Rose[1], Aly Fathy[1], Sumeet Kumar Gupta[2], Ahmedullah Aziz[1], *Senior Member, IEEE*

***Abstract*- Phase transition materials (PTM) have drawn significant attention in recent years due to their abrupt threshold switching characteristics and hysteretic behavior. Augmentation of the PTM with a transistor has been shown to provide enhanced selectivity (as high as ~$10^7$ for Ag/$HfO_2$/Pt) leading to unique circuit-level advantages. Previously, a unipolar PTM, Ag-$HfO_2$-Pt, was reported as a replacement for diodes due to its polarity-dependent high selectivity and hysteretic properties. It was shown to achieve ~50% higher DC output compared to a diode-based design in a *Cockcroft-Walton* multiplier circuit. In this paper, we take a deeper dive into this design. We augment two different PTMs (unipolar Ag-$HfO_2$-Pt and bipolar $VO_2$) with diode-connected MOSFETs to retain the benefits of hysteretic rectification. Our proposed hysteretic diodes (Hyperdiodes) exhibit a low forward voltage drop owing to their volatile hysteretic characteristics. However, augmenting a hysteretic PTM with a transistor brings an additional stability concern due to their complex interplay. Hence, we perform a comprehensive stability analysis for a range of threshold voltages ($-0.2 < V_{th} < 0.8$) and transistor sizes to ensure operational stability and to choose the most optimum design parameters. We then test a standalone Ag-$HfO_2$-Pt and an Ag-$HfO_2$-Pt-based Hyperdiode in two different types of voltage multipliers and report ~500 and ~20 times lower settling time, respectively. As the PTM possesses additional sources of variation, it is crucial to examine the performance benefits of the structure through an extensive variation analysis. We perform 3$\sigma$ Monte-Carlo variation analysis for a Cockcroft-Walton multiplier considering the nonidealities in the host transistor and the PTM. We observe that, Hyperdiode-based design achieves ~20% higher output voltage compared with the conventional designs within a fixed timeframe (200 $\mu$s).***

***Keywords*—Phase transition material, diode, hysteresis, voltage multiplier, Monte-Carlo, Hyperdiode.**

## I. Introduction

Transistor technology has undergone significant advancements in area scaling and power efficiency over the last few decades [1]. While it has transformed a multitude of technological paradigms, new challenges associated with the advancement continue to emerge [2]. As electronic industries are continually tackling these challenges through novel design approaches, extensive research efforts have been given to improve the transistor performance and applicability [3]. To achieve the best possible performance of an electronic device, it is also crucial to focus on optimizing other active components.

Manuscript received XXXXXXXXX; accepted XXXXXXXXX. Date of publication XXXXXXXXX; date of current version XXXXXXXXXX. This material is based in part on research sponsored by Air Force Research Laboratory under agreement FA8750-21-1-1018. This work was also supported in part by NSF. Award No: 2052780.

M. M. Islam, Shamiul Alam, Garrett S. Rose, Aly Fathy, and A. Aziz are with the Department of Electrical Engineering and Computer Science, University of Tennessee, Knoxville, TN, USA. (E-mail: mislam49@vols.utk.edu, salam10@vols.utk.edu, garose@utk.edu, afathy@utk.edu, aziz@utk.edu).

S. Gupta is with the School of Electrical and Computer Engineering, Purdue University, West Lafayette, IN, USA. (E-mail: guptask@purdue.edu)

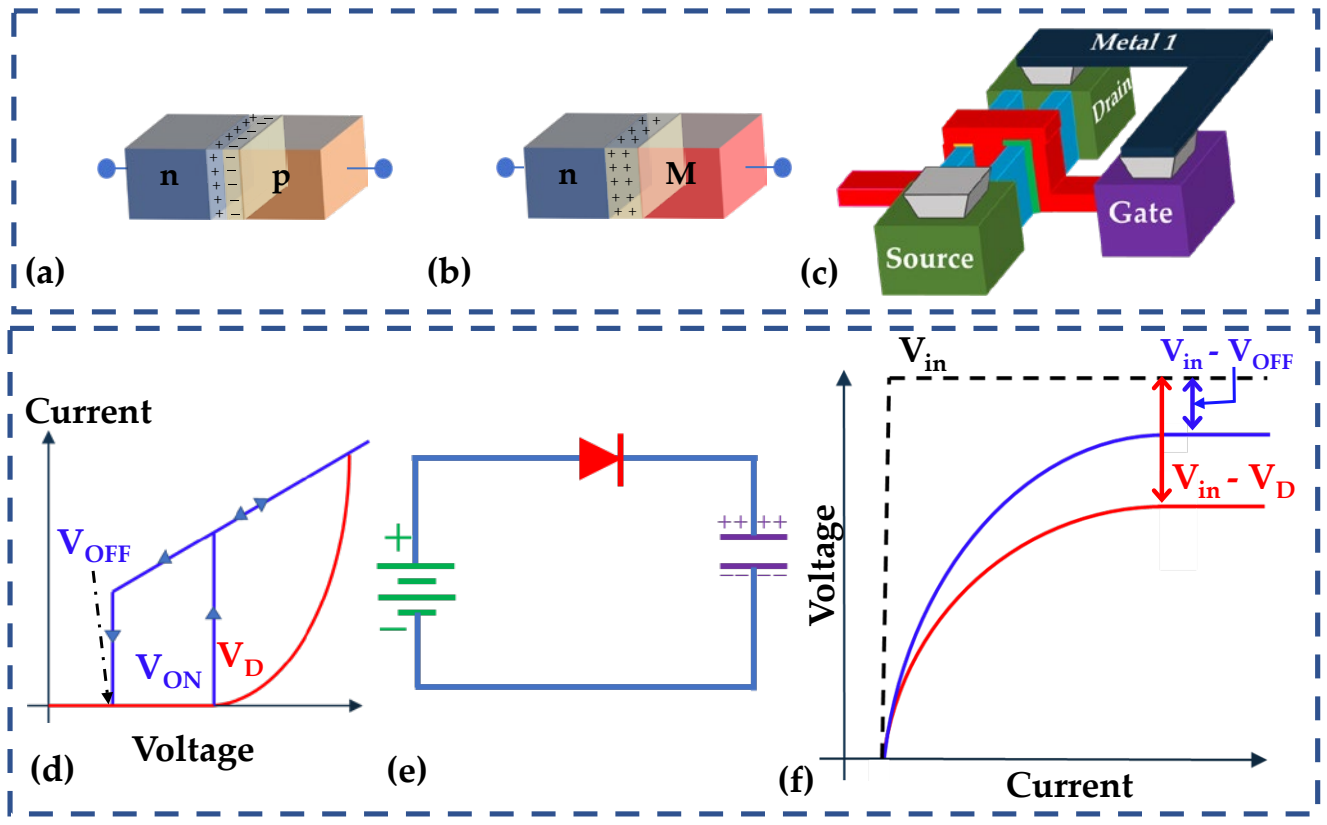

**Figure 1**: Physical structure of a **(a)** p-n diode **(b)** Schottky diode, and **(c)** diode connected FinFET. **(d)** I-V characteristics of the envisioned hysteretic diode (blue) with a conventional diode connected FinFET (red). **(e)** A simple circuit with a diode charging a capacitor. **(f)** Input voltage ($V_{in}$) and output voltage level for a hysteretic (blue) and a non-hysteretic diode (red).

A diode is one such indispensable active device component in electronic circuits playing crucial roles such as rectification, voltage regulation, signal modulation, and energy harvesting in mm-wave technology, etc [4], [5]. However, an on-chip diode faces several challenges such as high leakage current, forward voltage drop, reverse recovery transient, lower breakdown voltage, etc [6], [7]. Diode-connected FET (MOS-diode) incurs significant forward voltage drop, hurting their performance for low-voltage operation. While Schottky diodes exhibit low forward voltage drop and shorter reverse recovery time, they have drawbacks such as low reverse breakdown voltage and high leakage current. Consequently, for next-generation low-power electronics, a diode with low forward voltage drop and low leakage current is highly desired [8].

Recently, a unipolar PTM-based switch, Ag-$HfO_2$-Pt, has been proposed as a diode-replacement in a voltage multiplier circuit. Here, the unipolar PTM structure exhibits two discrete resistive states (insulating and metallic) and unique hysteretic characteristics.to achieve a low forward voltage drop while simultaneously maintaining a low off-current, thanks to its high insulating-state resistance [9]. However, the structure does not provide considerable flexibility and lacks practical tuning knobs as the diode dynamics is mainly dictated by the material properties. Here, we explore the possibility to combine a PTM with a diode-connected CMOS transistor to fully utilize the unique benefits of hysteretic rectification, without sacrificing design tunability. A wide variety of phase transition materials (PTMs), including both unipolar and bipolar types, have been extensively reported with diverse transition voltages and resistance levels [10]–[12]. Considering the inherent rectification property of the MOS-diode, it is reasonable to consider augmenting PTMs with both unipolar and bipolar characteristics to explore their implications in the augmented structure [13]–[15]. Hence, in addition to Ag-$HfO_2$-Pt as a unipolar PTM, we choose vanadium dioxide ($VO_2$) as a bipolar PTM in our analysis. $VO_2$ has been well-studied for decades and can be integrated with the current CMOS process technology through monolithic integration [12].

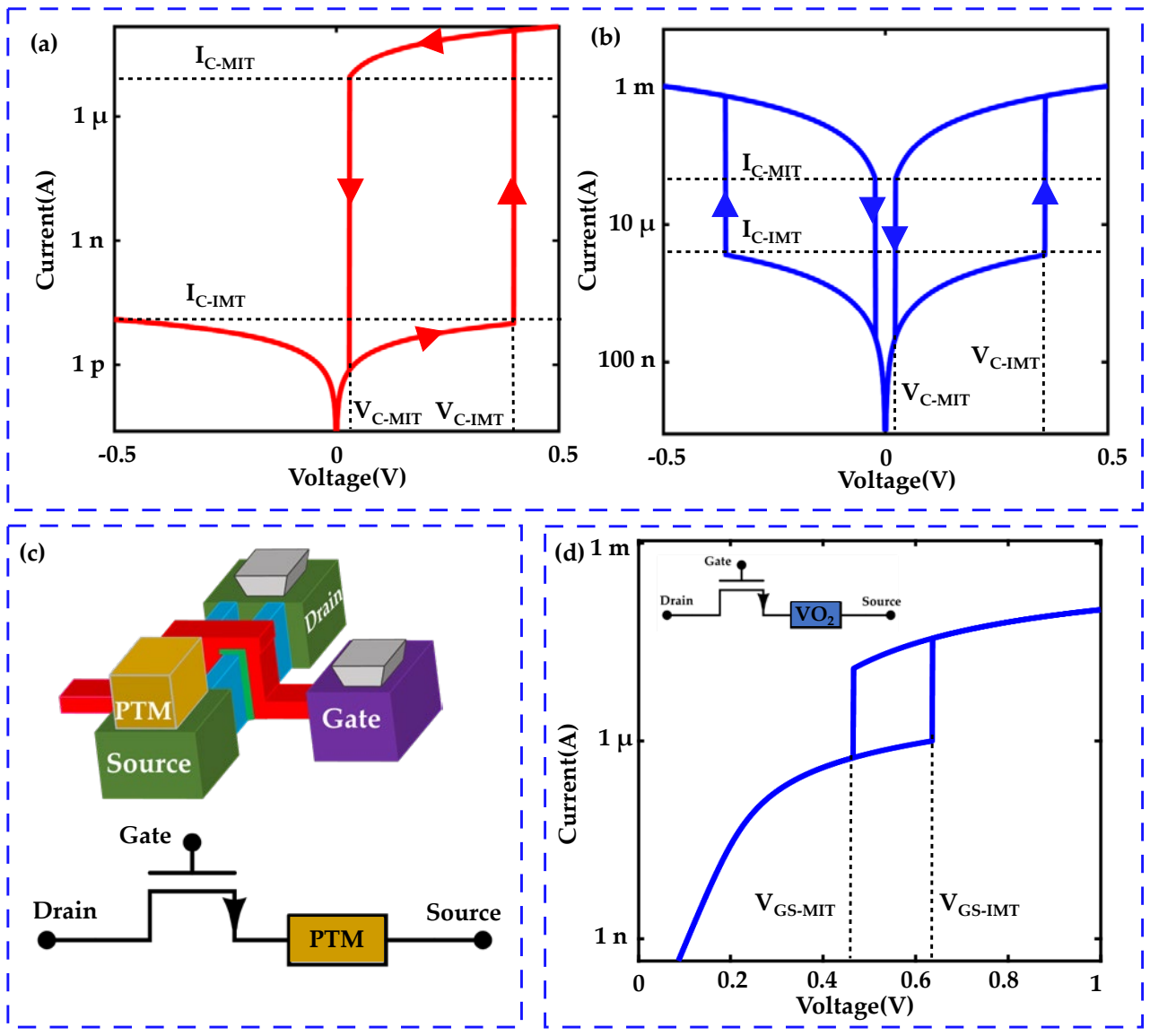

**Figure 2**: I-V characteristics of **(a)** Ag-$HfO_2$-Pt [36], and **(b)** $VO_2$ [37]. **(c)** Physical structure of a **(c)** Hyper-FET [13], and **(d)** its I-V characteristics.

The complex interplay between the MOS-diode and the PTM raises a concern for system stability due to the abrupt transition behaviour and hysteretic characteristics of the PTM [16]. For a PTM-augmented MOS-diode, it is of paramount interest to assess its stability within a relevant range of input voltages and inherent parameters. In this manuscript, we perform a comprehensive stability analysis for a PTM-augmented MOS-diode (Hyper-diode) for both unipolar and bipolar PTMs for varying threshold voltage and number of fins ($n_{fin}$). [17]. The insights provided by our analysis are highly useful for selecting suitable design parameters. In the realm of RF circuits and energy harvesting technology, the MOS-diode plays a critical role, with one of its important applications being the voltage multiplier.[18], [19]. We benchmark and evaluate the circuit-level performance of our proposed diode using two widely used voltage multiplier circuits. To assess the practicality of the novel rectifier platforms, it is necessary to consider the impact of variations. Process variation can significantly alter the threshold voltage and the channel current of the nanoscale transistors [20]. The PTM brings in additional sources of variation in our design in the resistance levels ($R_{INS}$ and $R_{MET}$) and the transition voltages ($V_{C-IMT}$ and $V_{C-MIT}$). Hence, it is crucial to perform a systematic variation analysis to ensure that considerable advantage is maintained even for the worst-case scenario We compare the variation tolerance of the standard MOS-diode and the proposed Hyper-diode through a 500-point Monte Carlo variation analysis [21].

The organization of this paper is as follows. Section II provides a brief overview of the PTM and introduces the concept of Hyperdiode. In section III, we perform a detailed stability analysis for $VO_2$ and Ag-$HfO_2$-Pt-based Hyperdiodes for different transistor sizes. In section IV, we show the performance comparison of a standalone PTM, Hyperdiode, and conventional MOS-diode in DCP and CW multiplier and evaluate the validity of our proposition. Finally, we test the degree of variation tolerance of our proposed Hyperdiode using the Monte Carlo variation analysis in section V.

## II. Phase Transition Material-Based Diode

### A. *Challenges Faced by Existing Diodes: An Overview*

A diode is a unidirectional active electronic component widely used in wireless communication, energy harvesting technology, and optical devices. Diodes can be classified into three major classes. a) p-n diode, b) Schottky diode c) MOS-diode (Fig. 1). However, conventional diodes often suffer from a common problem of forward voltage drop, which imposes limitations on the DC output voltage in rectifiers and voltage multiplier circuits.[18]. The forward voltage drop ($V_D$) is defined as the minimum voltage required to keep itself turned on. Although Schottky diode serves as a low voltage drop diode, it suffers from high reverse leakage current and low reverse breakdown voltage which sets significant limitations to its performance and reliability [22]. For MOS-diode, $V_D$ can be reduced by decreasing the threshold voltage ($V_{th}$) of the device. But the reduction of $V_{th}$ raises the leakage current ($I_{off}$) posing significant challenges for low power applications. Intuitively, both low $I_{off}$ and low $V_D$ are desired for a MOS-diode. Unfortunately, conventional diodes with typical I-V characteristics, where a diode turns on and off at the same voltage, cannot meet both requirements as a reduction of $V_D$ increases the $I_{off}$(Fig.1(d)). By independently reducing the turn-off voltage ($V_{OFF}$) of a diode, the voltage drop can be lowered without affecting the $I_{off}$ (as shown in Fig. 1(d)). The hysteretic diode can achieve both advantages independently. From a simple circuit analogy (Fig. 1(e)), it can be envisioned that a hysteretic diode wins over a conventional diode due to the low $V_D$ (= $V_{OFF}$) (Fig.1(f)).

### B. *Phase Transition Materials and Hyper-FET*

PTMs are special materials that exhibit an abrupt transition between insulating and metallic states triggered by several types of stimuli (*i.e.* electrical, optical, mechanical, and thermal) [23]. When the voltage across a PTM reaches a particular value ($V_{C-IMT}$), insulating to metallic transition occurs (Fig. 2(a,b)). The material further returns to its insulating state once the voltage across it goes below $V_{C-MIT}$ (Fig. 2(a,b)). The family of PTMs is quite rich with diverse transition voltages and resistivity levels. This gives the benefit of selecting and optimizing PTMs based on the specification required for the application. For example, both unipolar and bipolar PTMs have been reported and extensively used [13], [24]. We have incorporated a SPICE-compatible compact model for both the unipolar (Ag-$HfO_2$-Pt) and a bipolar ($VO_2$) PTM. For the host FinFET, the predictive technology model has been used. In Ag-$HfO_2$-Pt, the Ag filament is formed in the interstitial sites of $HfO_2$ (Fig.3(a)) [24]. Due to the inert nature of the Pt electrode, no filament formation occurs when a voltage is applied across the Pt terminal. This asymmetry makes Ag-$HfO_2$-Pt unipolar.

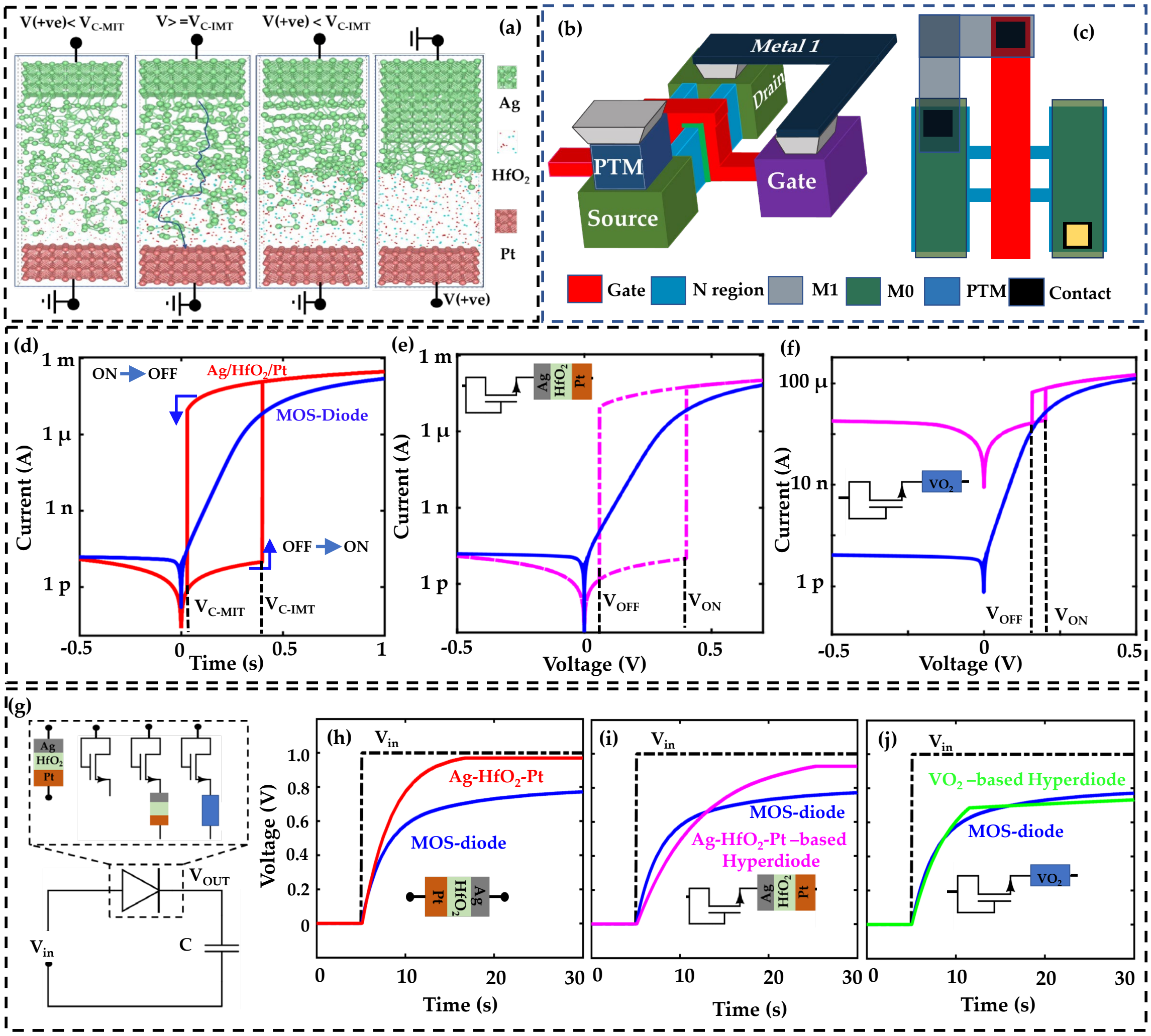

**Figure 3**: **(a)** Physical mechanism of unipolar switching of Ag-HfO2-Pt. Structure [40] and layout of **(b)** Structure and **(d)** layout of a Hyperdiode. **(d)** I-V characteristics of a MOS-diode (blue) and Ag-HfO2-Pt (red), designed to possess similar on-current and turn-on voltage ($V_D \approx V_{C\text{-}IMT}$). I-V characteristics of a **(e)** Ag-HfO2-Pt-based Hyperdiode and **(f)** $VO_2$-based Hyperdiode side by side with the matched diode. **(g)** Circuit for charging a capacitor by four different versions of diodes. Comparison of $V_{OUT}$ for diode-connected FET and **(h)** standalone Ag-HfO2-Pt, **(i)** Ag-HfO2-Pt-based Hyperdiode and (j) $VO_2$-based Hyperdiode.

$VO_2$ is a bipolar PTM where a strong electron correlation drives the phase transition mechanism.

Among the wide variety of materials reported so far, $VO_2$ has been used with a MOSFET device giving rise to the hysteretic I-V characteristics [12]. This hybrid system has been referred to as the hybrid phase transition FET (Hyper-FET) where the simple augmentation of PTM with conventional MOSFET has overcome the fundamental Boltzmann limit (60 mV/dec) (Fig. 2(c)). The Hyper-FET demonstrates hysteresis in its I-V characteristics (as shown in Fig. 2(d)) enhancing the performance of various digital and analogue devices and circuits.[25]–[27]. Since the Hyper-FET exhibits hysteretic behavior due to the augmented PTM, intuitively the Hyper-diode is supposed to exhibit rectification characteristics with unidirectional hysteretic behavior.

## C. Hyperdiode

Augmenting a unipolar PTM with a FET device provides additional high resistance during the off state of the host transistor reducing $I_{off}$ substantially without introducing any additional area penalty (Fig.3(b,c)). In our model, Ag-$HfO_2$-Pt has very high off resistance ($R_{OFF}$ = 40 GΩ) for which it dominates the overall characteristics during off state. During on-state, it adds a resistance that suppresses the on-state current. However, by reducing the transistor threshold voltage, it can considerably compensate the overall $I_{on}$ of the device. This way, it is possible to match the $I_{on}$ of the host transistor lowering $I_{off}$ substantially. Additionally, the hysteretic characteristics of the device result in the improvement of the forward voltage drop problem due to the very low value of $V_{C\text{-}IMT}$ (Fig. 3(d)).

While unipolar Ag-$HfO_2$-Pt inherently works as a rectifier, bidirectional $VO_2$ does not exhibit rectification property as it undergoes insulator to metal transition (IMT) from both directions. However, it can be augmented with a MOS-diode to achieve rectification. When, a voltage is applied from opposite direction, $VO_2$ does not offer high resistance owing to its bipolarity. But it still reduces $I_{off}$ substantially due to its insulating state resistance. Moreover, it introduces hysteresis in the *I-V* characteristics of the device lowering the forward voltage drop. Fig. 3(d) shows the *I-V* characteristics of a standalone Ag-$HfO_2$-Pt and the host transistor. Fig. 3(e) shows the *I-V* characteristics of the Ag-$HfO_2$-Pt-based Hyperdiode. Here, the host transistor is engineered so that the $I_{on}$ of both devices match. The off-current ($I_{off}$) decreases by several orders of magnitude, thanks to the high insulating-state resistance ($R_{INS}$ = 40 GΩ) of the augmented PTM. Fig. 3(f) shows the *I-V* characteristics of a $VO_2$-based Hyperdiode. Here, the host transistor is engineered to match the on-current of the MOS-diode. For both of these cases, the turn-off voltage ($V_{OFF}$) is determined by the metal-to-insulator transition voltage ($V_{C\text{-}MIT}$) of the PTM which is much lower than the threshold voltage of the MOS-diode. The advantage of using both Hyperdiodes are shown in Fig.3(g-j). It is evident that, using a standalone Ag-$HfO_2$-Pt substantially improves the speed of charging and substantially reduces the forward voltage drop ($V_D$). Ag-$HfO_2$-Pt-based Hyperdiode shows less improvement compared to the standalone Ag-$HfO_2$-Pt but still manages to achieve faster charging with a lower $V_D$ compared to the MOS-diode. Additionally, the host transistor makes the Hyperdiode CMOS-compatible and thus fabrication-friendly. The $VO_2$-based Hyperdiode suffers from the same $V_D$ due to the narrow hysteresis in its characteristics.

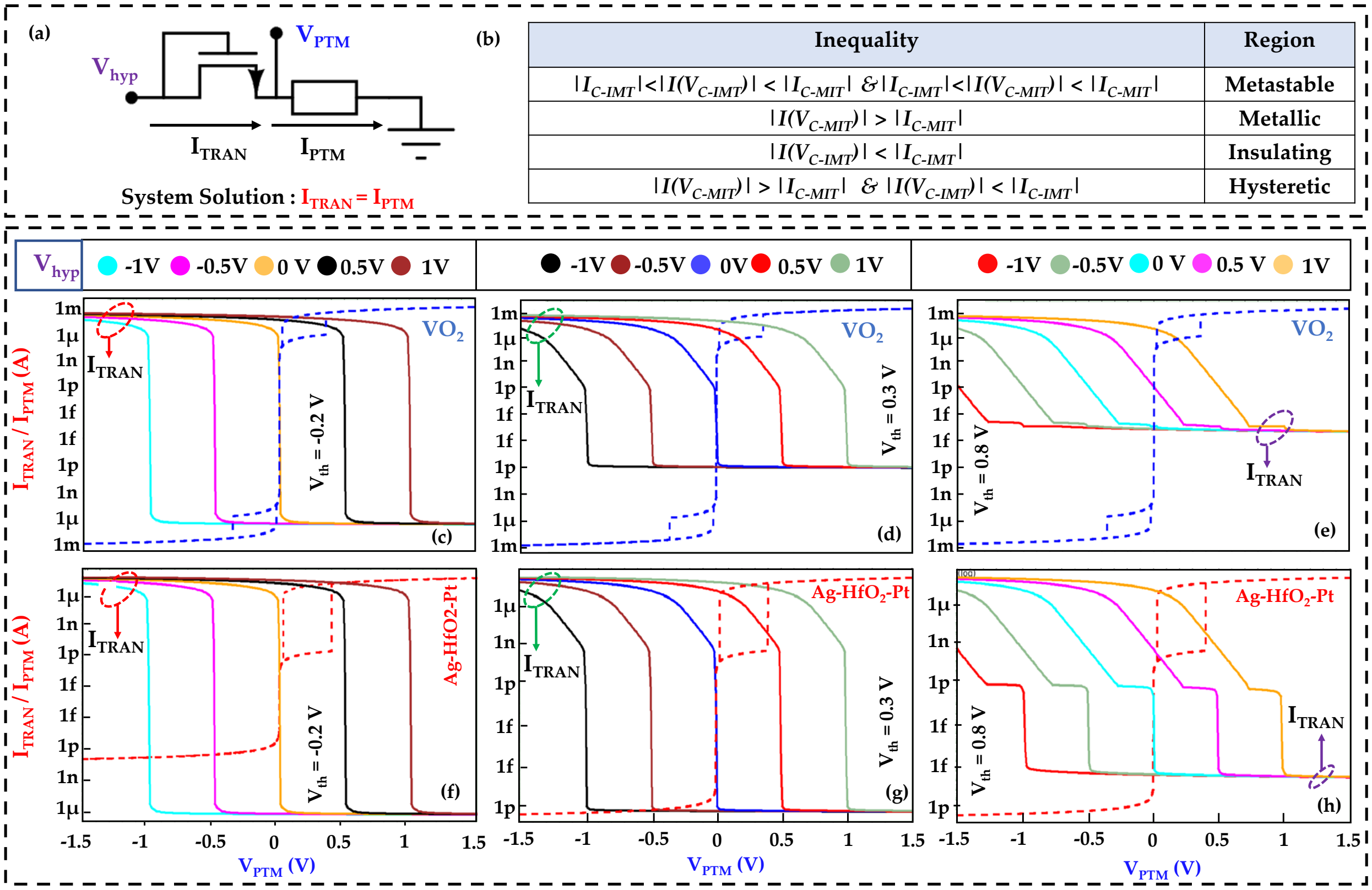

**Figure 4**: **(a)** Circuit for stability analysis of the system. **(b)** Inequality criteria for different region of stability. I-V characteristics of a standalone $VO_2$ and diode-connected FET for different source voltage (-1 V, -0.5 V, 0 V, 0.5 V,1 V) with **(c)** $V_{th}$ = -0.2 V **(d)** $V_{th}$ = 0.3 V and **(e)** $V_{th}$ = 0.8 V. I-V characteristics of a standalone Ag-HfO2-Pt device and diode connected FET for different source voltage (-1 V, -0.5 V, 0 V, 0.5 V,1 V) with **(f)** $V_{th}$ = -0.2 V **(g)** $V_{th}$ = 0.3 V and **(h)** $V_{th}$ = 0.8 V.

In contrast to the smooth and continuous current-voltage relationship of the MOS-diode, a PTM exhibits a sharp and hysteretic characteristic, allowing only certain current values to pass through it. As a result, when connected with a FET in series, concerns arise about the stability of the system due to the shared current. In the following section, we will conduct a thorough stability analysis for the whole structure.

## III. Stability Analysis

To determine the stability of a system consisting of two components in series, one can determine the stable current value that is shared between both components. This is executed by plotting the current across each component against the voltage at the same node, and then identifying the intersection point of both characteristic curves. This intersection point provides the solution of the system, which can be used to analyze its stability states. Here, we plot the *I-V* characteristics of the MOS-diode and the PTM with respect to the common node voltage ($V_{PTM}$) as shown in Fig.4(a). The point where the two characteristic curves intersect indicates the shared solution of the system, in which $I_{TRAN} = I_{PTM}$. When the curves meet at a steady current value of PTM (either metallic or insulating), the system is stable for that specific voltage ($V_{hyp}$), and PTM becomes stable (either insulating or metallic). Conversely, when the curves intersect at the metastable points only, the system stability fails. Again, if the *I-V* curve of the MOS-diode intersects with both the stable state's current values (insulating and metallic), then the PTM can stabilize to either of the states depending on the initial state of the PTM. These bi-stability points indicate the hysteretic characteristics of the system. For the hysteretic diode to be beneficial, it must maintain stability while also having a hysteretic zone within the range of operating voltages. Fig. 4(b) summarizes the stability criteria for the Hyperdiode structure.

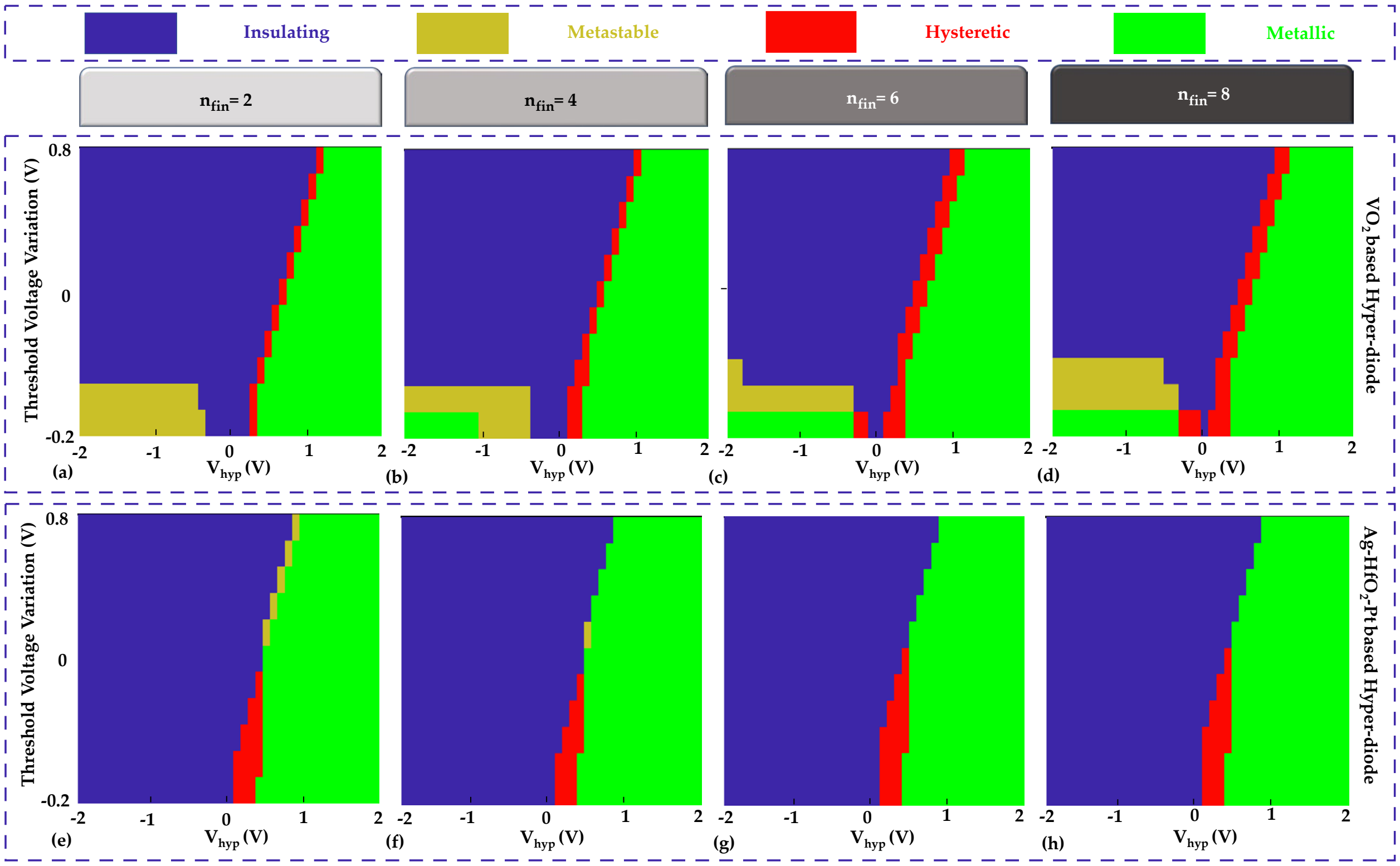

**Figure 5:** Colormap describing the stability for the circuit in fig. 4(a). Stability for different voltage across the $VO_2$-based Hyperdiode (X-axis) and different threshold voltage for the transistor (Y-axis) with **(a)** $n_{fin} = 2$, **(b)** $n_{fin} = 4$, **(c)** $n_{fin} = 6$ and **(d)** $n_{fin} = 8$. Stability for different voltage across the Ag-$HfO_2$-Pt-based Hyperdiode and different $V_{th}$ for the transistor with **(e)** $n_{fin} = 2$, **(f)** $n_{fin} = 4$, **(g)** $n_{fin} = 6$ and **(h)** $n_{fin} = 8$.

The host transistor's threshold voltage ($V_{th}$) is a tunable parameter that can be precisely adjusted to meet the diode's operational requirements. In case the system fails to stabilize at the nominal $V_{th}$, we can vary the threshold voltage to approach a stable solution. Additionally, to make our analysis more comprehensive, we have also varied the number of fins($n_{fin}$). The mechanism of our stability analysis is depicted in Fig.4. We have swept the $V_{th}$ of the MOS-diode at every 0.1 V interval for $-0.2V < V_{th} < 0.8V$ and $V_{hyp}$ is varied at a step size of 0.1 V for $-2V < V_{hyp} < 2V$.

The bidirectional hysteretic behavior of a bipolar PTM ($VO_2$) creates a stability concern, even when a negative voltage is applied across the Hyperdiode, as evident from Fig. 4(c). For the Ag-$HfO_2$-Pt-based Hyperdiode, the system's instability is caused by large hysteresis at a specific voltage. We summarize the result of our analysis in Fig. 5 through a colormap. In the case of $VO_2$, the PTM stabilizes in either the insulating or metallic resistive states when $V_{th} > -0.05V$ and $n_{fin}$ is between 2 to 4. For $V_{th} > 0$ V, stability is maintained for $n_{fin} > 6$. Based on our analysis, $V_{th} = 0$ V with $n_{fin} = 8$ provides the most optimal characteristic for a $VO_2$-based Hyperdiode. On the other hand, for the Ag-$HfO_2$-Pt-based Hyperdiode, the unstable region can be eliminated by increasing $n_{fin}$. However, to take advantage of the hysteretic behavior in lower threshold voltage, selecting the lowest $n_{fin}$ while considering the area advantage is ideal. Therefore, the most optimal parameter selection for the Ag-$HfO_2$-Pt-based hyperdiode would be $V_{th} = -0.2$ V and $n_{fin} = 2$.

## IV. Hyperdiode-Based Voltage Multiplier

Voltage multipliers have become increasingly important in a wide range of electronic circuit applications. They are particularly crucial for energy-autonomous and self-powering devices such as smart sensors and Body Area Networks (BANs) that utilize energy harvesting circuits to achieve higher voltage.[28]. Typically, energy harvesting circuits take a small transient AC signal as an input to create DC voltage output with higher values. The CW voltage multiplier is one such multiplier circuit widely used in energy harvesting technology. [29]. The DCP is a voltage multiplier topology commonly employed in integrated circuits to raise a low-voltage battery supply to the voltage level necessary for proper IC functioning. Additionally, it is widely used in energy harvesting applications such as photovoltaic cells and thermoelectric generators, where the voltage generated by these sources is often inadequate for other circuit components. In such cases, a DC-to-DC boost converter is needed, and the DCP can be utilized to perform this function.[30]. We have incorporated the standalone Ag-$HfO_2$-Pt and Ag-$HfO_2$-Pt–based Hyperdiode in these multipliers.

### A. Dickson Charge Pump (DCP)

The basic circuit diagram of a DCP voltage multiplier is presented in Fig. 6(a). Here, an alternating capacitor stage network is powered by two non-overlapping clock signals with matching periods. To achieve maximum efficiency, a DCP is designed to transfer charge from one capacitor stage to the other during the subsequent clock pulses. An ideal n-stage DCP has an output of $(n+1)\times V_{hyp}-n\times V_T$, where $V_T$ is the threshold voltage of the switching device. The voltage level transferred to the subsequent capacitors is restricted by the forward voltage drop of the diode. Despite the widespread use of Schottky diodes in DCP circuits due to their low voltage drop, their performance remains suboptimal due to the high leakage current and high reverse recovery time. It is believed that diodes with hysteresis in their current-voltage (*I-V*) characteristics could offer faster response times. In our study, we utilize two non-overlapping clock signals with a frequency of 1 MHz with a DC voltage input of 1V. Fig. 6(b) illustrates the final stage output of a four-stage DCP for a MOS-diode, standalone Ag-$HfO_2$-Pt, and Ag-$HfO_2$-Pt-based Hyperdiode. To evaluate their performance, we selected a 200 µs timeframe for our analysis. Our results indicate that the Ag-$HfO_2$-Pt diode exhibits the fastest settling speed, while the proposed Hyperdiode outperforms the conventional MOS-diode by a significant margin of 0.4 V within the timeframe.

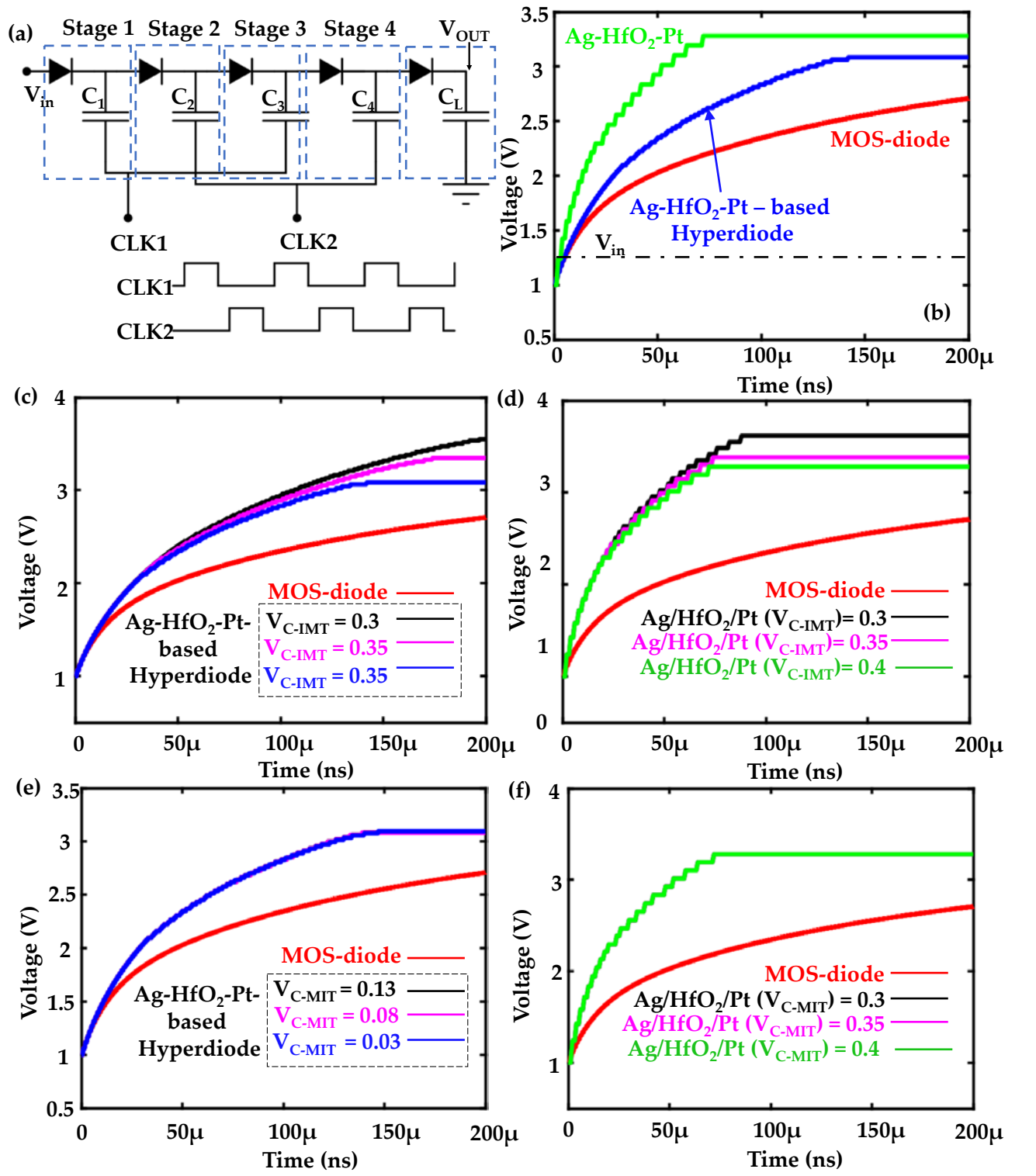

**Figure 6: (a)** Basic Dickson Charge Pump (DCP) voltage multiplier. **(b)** output of a 4-stage DCP with diode connected FET (red), standalone PTM (green) and Hyperdiode (blue) in response to a DC input, $V_{hyp}$ (black). Comparison of $V_{OUT}$ of a conventional diode based DCP and a Hyperdiode with **(c)** three different $V_{C\text{-}IMT}$ and (e) three different $V_{C\text{-}MIT}$. Comparison of the $V_{OUT}$ of a conventional diode-based DCP and a standalone Ag-$HfO_2$-Pt with **(d)** three different $V_{C\text{-}IMT}$ and (f) three different $V_{C\text{-}MIT}$.

For future improvement of a DCP voltage multiplier, we have varied the $V_{C\text{-}IMT}$, and $V_{C\text{-}MIT}$ to observe the transient response. Fig. 6(c-d) illustrates the output voltage comparison for three different $V_{C\text{-}IMT}$ (0.3,0.35,0.4). At lower $V_{C\text{-}IMT}$, the diode turns on earlier which results in faster settling time and higher output voltage. On the other hand, varying $V_{C\text{-}MIT}$ hardly affects the output response. These findings suggest that incorporating a PTM with lower VC-IMT could significantly enhance the response of a DCP voltage multiplier, offering scope for future improvement.

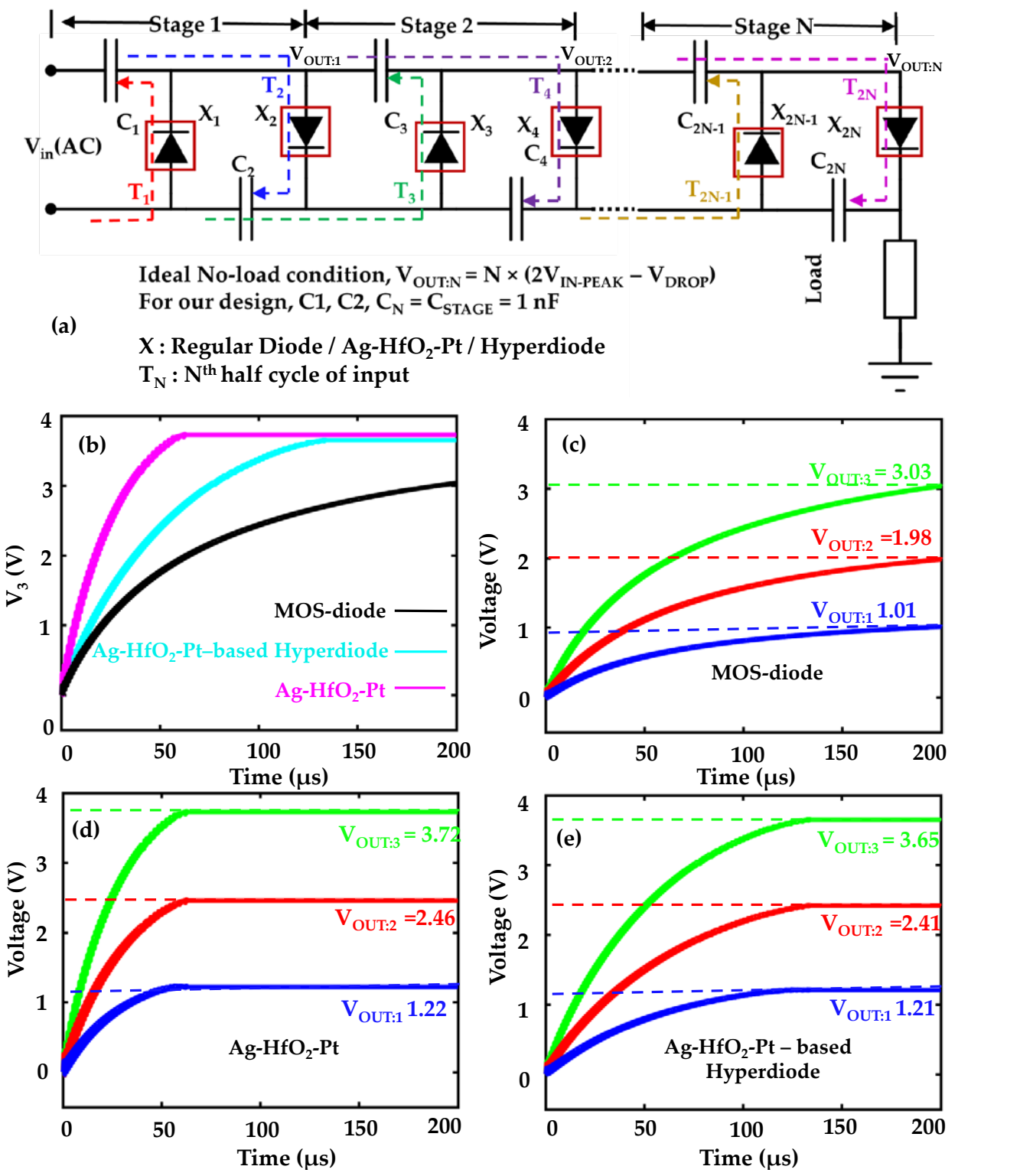

**Figure 7: (a)** A generalized circuit diagram of a Cockcroft Walton (CW) multiplier [11]. **(b)** Output voltage for a 3-stage CW multiplier circuit for a diode-connected FET, standalone Ag-$HfO_2$-Pt and Ag-$HfO_2$-Pt–based Hyperdiode. **(c)-(e)** The output voltages at different stages for a 3-stage CW multiplier for diode connected FET, standalone Ag-$HfO_2$-Pt and Ag-$HfO_2$-Pt-based Hyperdiode.

### *B. Cockcroft-Walton (CW) voltage multiplier*

The basic Cockcroft Walton (CW) voltage multiplier is depicted in Fig. 7(a). The diode-capacitor network accumulates different levels of DC voltage from a small AC input signal over subsequent clock cycles at the output. In the context of energy-harvesting operation, a faster response is desired. To compare the performance, we have chosen a timeframe of 200 $\mu$s. During no-load condition, the $N^{th}$ stage CW multiplier produces an output, $V_{OUT:\,N} = 2\times N\times V_P - N\times V_{DROP}$. Fig. 7(b) shows the value of $V_{OUT}$ for a 3-stage CW multiplier. The Standalone Ag-$HfO_2$-Pt has a 23% larger $V_{OUT}$ than the MOS-diode, while Ag-$HfO_2$-Pt-based Hyperdiode shows a 20% higher $V_{OUT}$. Fig. 7(c-e), shows the side-by-side comparison between different stages of a CW multiplier. The settling times for the Ag-$HfO_2$-Pt device, Hyperdiode, and MOS diode are estimated as 5µs, 120µs, and 2.5ms, respectively. Compared to the hysteretic diodes being considered (Ag-$HfO_2$-Pt and Hyperdiode), a MOS-diode has a much longer settling time, taking several milliseconds. This is due to the sharp transition behavior and lower turn-off voltage of the hysteretic diodes, which allows them to remain turned on for a higher duration. Therefore, we observe a faster charging in these cases. Once the voltage levels at different stages reach certain values, the drop across the individual hysteretic diodes reaches below $V_{C\text{-}MIT}$ and the voltage level settles at its maximum value. Although the voltage levels in the MOS-diode can eventually exceed those of the hysteretic diodes over time, its slower settling time makes it unsuitable for applications that require fast response.

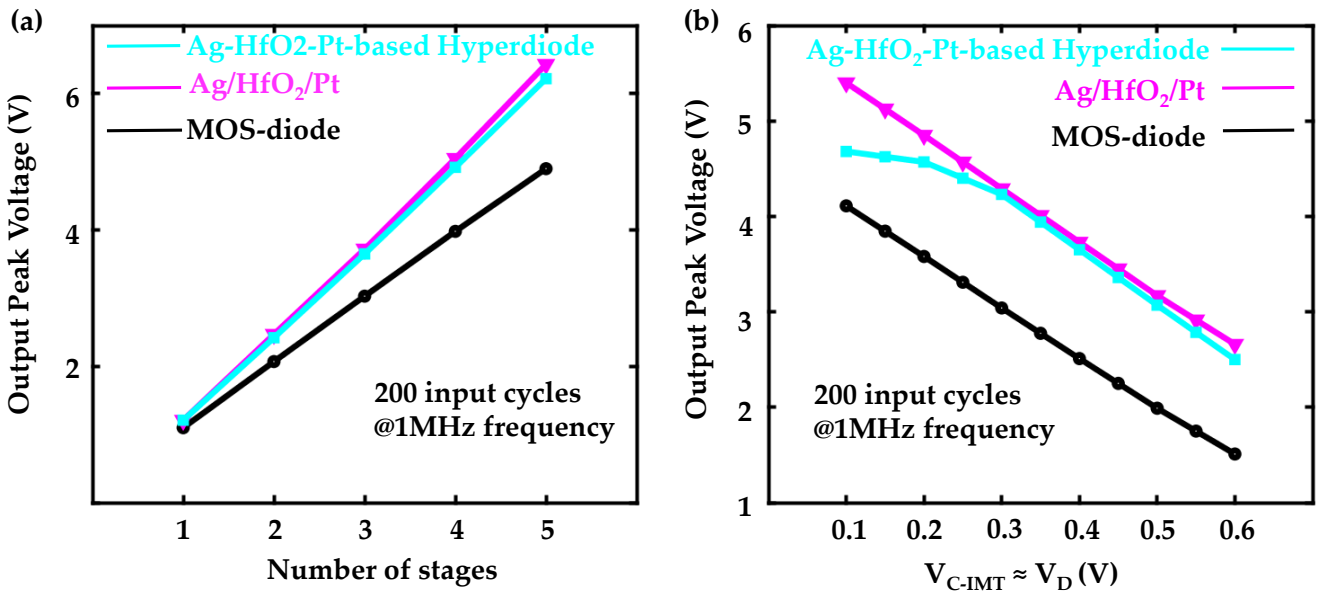

**Figure 8: (a)** Comparison of output peak voltage for different total number of stages for Ag-$HfO_2$-Pt (purple), Hyperdiode (blue) and MOS-diode (black). **(b)** Comparison of output peak voltage for different switching voltage ($V_{C\text{-}IMT}$ ~ $V_D$) for Ag-$HfO_2$-Pt (purple), Hyperdiode (blue) and MOS-diode (black).

We have extended our analysis by investigating the impact of the number of stages of the CW multiplier on the output voltage level. With an increase in the number of stages, the MOS-diode takes longer to settle down compared to the hysteretic diodes, resulting in a greater voltage difference between them. (Fig. 8(a)). We have examined how the output voltage level of the CW multiplier is affected by varying the $V_{C\text{-}IMT}$. To ensure a fair comparison, we have varied $V_{th}$ of the diode-connected FET accordingly. With lower $V_{C\text{-}IMT}$ ($\approx V_D$), we get higher output voltage levels than the standalone MOS-diode (Fig. 8(b)). However, when $V_{C\text{-}IMT}$ is extremely low, the Ag-$HfO_2$-Pt loses its effectiveness in the Hyperdiode, causing it to approach the same value as the MOS-diode. Hence, we see the Hyperdiode approaching the MOS-diode at a very low $V_{C\text{-}IMT}$. Overall, the augmentation of a PTM with the MOS- diode decreases the output voltage level slightly, but in return, we achieve manufacturing compatibility giving the proposed Hyperdiode a practical advantage.

## V. Variation Analysis

The slight dimensional variation of a PTM can alter the transition voltages ($V_{C\text{-}IMT}$ and $V_{C\text{-}MIT}$) and the level of resistivity ($R_{INS}$ and $R_{MET}$). Additionally, it is important to consider the variation of the threshold voltage of a MOSFET to ensure reliable performance. To determine if our hysteretic diodes offer an advantage over conventional diode-connected FETs in all scenarios, we have examined the impact of variations in three key parameters: insulator to metal transition voltage ($V_{C\text{-}IMT}$), insulating state resistance ($R_{INS}$), and transistor threshold voltage ($V_{th}$). These parameters have significant implications on their performance in device and circuit level. We have disregarded the possible variation of $V_{C\text{-}MIT}$ as it has minimal impact on the performance of the CW multiplier.

We perform Monte Carlo variation analysis for both the Hyperdiode and the standalone Ag-$HfO_2$-Pt device. Here, we impose a Gaussian distribution of $3\sigma$ -variation in the parameters under consideration. The specification of the variation is given in Fig. 9(b). The nominal value of $V_{th}$ is chosen to be -0.2V and kept at the same value for the host transistor of the Hyperdiode. We have estimated the output voltage of a 3-stage CW multiplier at the time instance of 200 $\mu$s. Our proposed Hyperdiode has an additional degree of variation, leading to a higher voltage spread in the observed distribution compared to the standalone Ag-$HfO_2$-Pt device. Despite a higher spread in both cases, the standalone Ag-$HfO_2$-Pt and Ag-$HfO_2$-Pt-based Hyperdiode offer higher $V_{OUT}$ even in the worst-case variation.

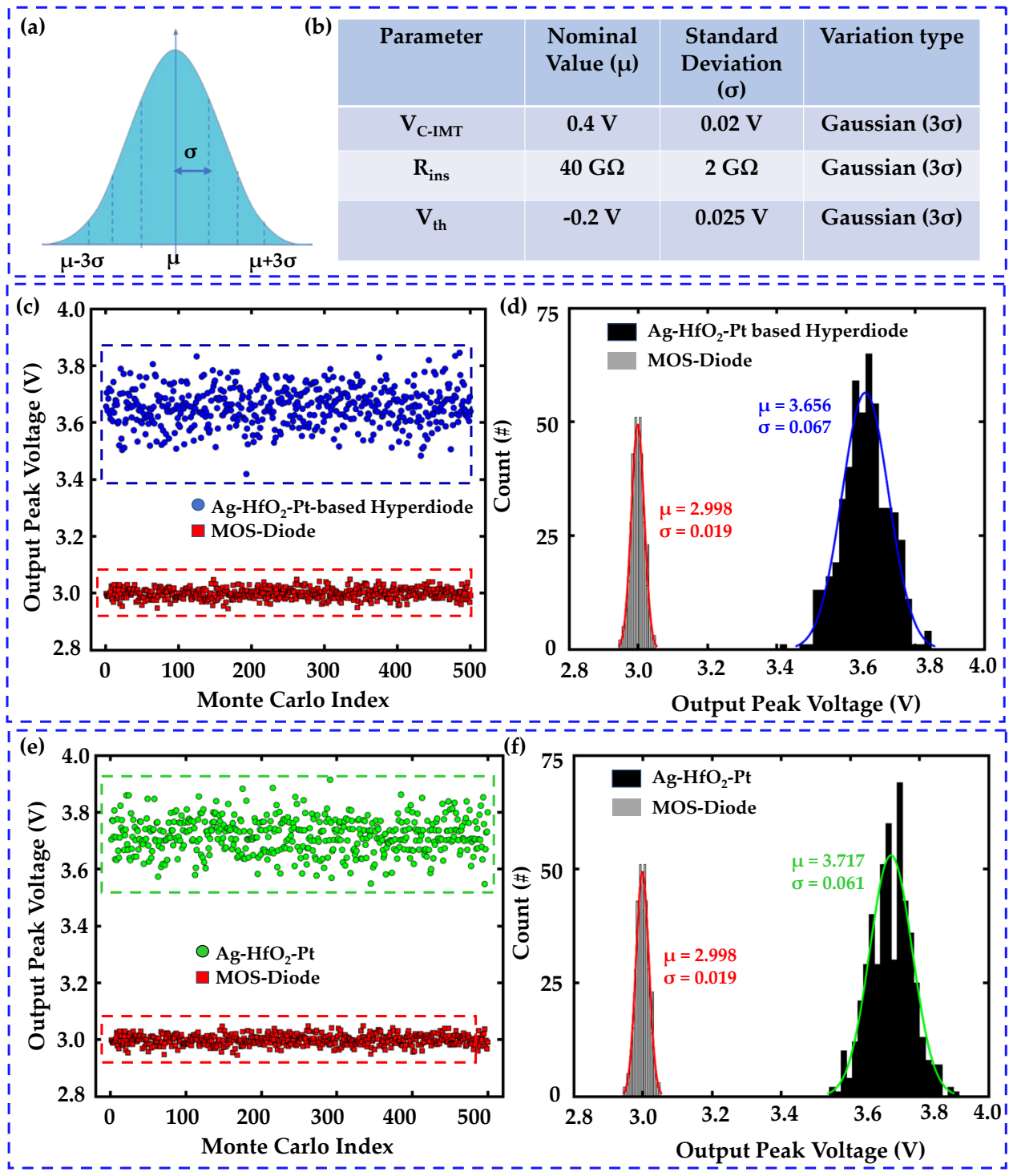

**Figure 9: (a)** A random gaussian distribution **(b)** Parameter specification for the variation analysis. **(c)** Scatter diagram and **(d)** histogram for output peak voltage for Ag-$HfO_2$-Pt-based Hyperdiode and MOS-diode. **(e)** Scatter diagram and **(f)** histogram for output peak voltage for a standalone Ag-$HfO_2$-Pt and MOS-diode.

## VI. Discussion

While the PTM-based Hyperdiode shows promise, there are practical considerations that need addressing. The first concern is the repetitive switching capability of the materials. Ag-$HfO_2$-Pt has been reported to switch reliably up to $10^8$ cycles, making it suitable for voltage multiplier circuits compared to $VO_2$, which is limited to $10^3$ cycles [24]. Besides, the switching time of the PTM is a critical factor that can significantly impact its overall performance. The filamentation process in Ag-$HfO_2$-Pt takes around 1ns [31], which is negligible compared to the operation timescale of voltage multipliers ($\mu s$).Temperature stability is another important aspect. Ag-$HfO_2$-Pt can withstand temperatures up to 90°C, whereas $VO_2$ has a lower insulator to metal switching temperature ($T_{IMT}$) of 68°C.[24]. While modern CMOS devices have been shown to operate at much higher temperatures, the thermal stability of our proposed Hyperdiode is limited by the temperature stability of Ag-$HfO_2$-Pt, which can withstand up to 90°C [32]. Further efforts are needed to improve the range of operating temperature.

In the context of RF energy harvesting, the Ag-$HfO_2$-Pt-based Hyperdiode shows promise as it offers a low forward voltage drop (< 0.1 V) and significantly lower off-current (~10 pA) compared to Schottky diodes, indicating superior speed and energy efficiency. Besides, the PTMs used in the proposed Hyperdiode do not exhibit reverse breakdown voltage or reverse recovery current, unlike conventional diodes, which is a favourable characteristic as the reverse breakdown voltage is a crucial factor that limits the operating voltage range of conventional diodes [33].

## VII. Conclusion

Our study utilizes both unipolar and bipolar PTM for our proposed hysteretic Hyperdiode. Through stability analysis, we determine the most optimal performance parameters for both $VO_2$-based Hyperdiode and Ag-$HfO_2$-Pt-based Hyperdiode. Although, both $VO_2$-based Hyperdiode and Ag-$HfO_2$-Pt-based Hyperdiode provide expected hysteretic behavior, $VO_2$-based Hyperdiode has a much higher off current than conventional MOS-diode owing to the adjusted low $V_{th}$ for stability. So, we proceed with the rest of our study on standalone Ag-$HfO_2$-Pt and Ag-$HfO_2$-Pt-based Hyperdiode. We incorporate this hysteretic diode in two voltage multiplier circuits and observe ~21% and ~15% higher voltage output compared to MOS-diode in a 4-stage DCP and 23% and 20% higher output voltage in a 3-stage CW multiplier (both at 200 µs). Finally, we perform a Monte-Carlo variation analysis for a 3 stage CW-multiplier by imposing a gaussian variation to three different chosen parameters. In addition, Monte-Carlo variation analysis is performed for a 3-stage CW multiplier by applying gaussian variation to three different parameters. Our results demonstrate that our proposed Hyperdiodes continue to outperform MOS-diode even in the worst-case variation by a significant margin.